\documentclass[11pt]{article}
\usepackage{amssymb}
\newcommand{\dshv}{
\begin{picture}(52,10)
\put(15,3){\line(-5,2){10}}
\put(15,3){\line(-5,-2){10}}
\multiput(15,3)(22,0){2}{\circle*{2}}
\multiput(15,3)(6,0){4}{\line(1,0){4}}
\put(37,3){\line(5,2){10}}
\put(37,3){\line(5,-2){10}}
\end{picture}  }
\newcommand{\vvertice}{
\begin{picture}(50,10)
\linethickness{0.25pt}
\put(15,3){\line(-5,2){10}}
\put(15,3){\line(-5,-2){10}}
\multiput(15,3)(20,0){2}{\circle*{2}}
\bezier{100}(15,3)(25,11)(35,3)
\bezier{100}(15,3)(25,-5)(35,3)
\put(35,3){\line(5,2){10}}
\put(35,3){\line(5,-2){10}}
\end{picture}  }
\newcommand{\six}{
\begin{picture}(40,10)
\put(15,3){\line(-5,2){10}}
\put(15,3){\line(-5,-2){10}}
\multiput(14,3)(12,0){2}{\circle*{2}}
\put(5,3){\line(1,0){30}}
\put(25,3){\line(5,2){10}}
\put(25,3){\line(5,-2){10}}
\end{picture}  }
\newcommand{\vloop}{
\begin{picture}(30,12)
\put(15,1){\line(-1,0){10}}
\put(15,1){\circle*{3}}
\put(15,6.5){\circle{10}}
\put(15,1){\line(1,0){10}}
\end{picture}  }
\newcommand{\sunset}{
\begin{picture}(42,10)
\multiput(15,4)(12,0){2}{\circle*{2}}
\put(5,4){\line(1,0){32}}
\put(21,4){\circle{12}}
\end{picture}  }
\newcommand{\dshvv}{
\begin{picture}(72,10)
\put(15,3){\line(-5,2){10}}
\put(15,3){\line(-5,-2){10}}
\multiput(15,3)(16,0){2}{\circle*{2}}
\multiput(41,3)(16,0){2}{\circle*{2}}
\multiput(15,3)(6,0){3}{\line(1,0){4}}
\multiput(41,3)(6,0){3}{\line(1,0){4}}
\put(36,3){\circle{10}}
\put(57,3){\line(5,2){10}}
\put(57,3){\line(5,-2){10}}
\end{picture}  }
\newcommand{\dshvvv}{
\begin{picture}(54,16)
\put(15,6){\line(-5,2){10}}
\put(15,6){\line(-5,-2){10}}
\multiput(15,6)(16,0){2}{\circle*{2}}
\multiput(39,-2)(0,16){2}{\circle*{2}}
\multiput(15,6)(6,0){3}{\line(1,0){4}}
\multiput(39,-2)(0,6){3}{\line(0,1){4}}
\put(39,6){\oval(16,16)[l]}
\put(39,-2){\line(1,0){10}}
\put(39,14){\line(1,0){10}}
\end{picture}  }
\newcommand{\dshvvvv}{
\begin{picture}(46,16)
\multiput(5,-2)(0,16){2}{\line(1,0){36}}
\multiput(15,-2)(0,6){3}{\line(0,1){4}}
\multiput(31,-2)(0,6){3}{\line(0,1){4}}
\multiput(15,-2)(0,16){2}{\circle*{2}}
\multiput(31,-2)(0,16){2}{\circle*{2}}
\end{picture}  }

\begin{document}

\title{
\textbf{Some aspects of the nonperturbative renormalization of the
$\varphi^4$ model}
}

\author{J. Kaupu\v{z}s
\thanks{E--mail: \texttt{kaupuzs@latnet.lv}} \\
Institute of Mathematics and Computer Science, University of Latvia\\
Rai\c{n}a bulv\={a}ris 29, LV--1459 Riga, Latvia}

\date{\today}

\maketitle

\begin{abstract}
A nonperturbative renormalization of the $\varphi^4$ model is considered.
First we integrate out only a single pair of conjugated modes
with wave vectors $\pm {\bf q}$. Then we are looking for the RG
equation which would describe the transformation of the Hamiltonian
under the integration over a shell
$\Lambda - d\Lambda < k < \Lambda$, where $d \Lambda \to 0$.
We show that the known Wegner--Houghton equation is consistent with the assumption of 
a simple superposition of the integration results for $\pm {\bf q}$.
The renormalized action can be expanded in powers of
the $\varphi^4$ coupling constant $u$ in the high temperature phase
at $u \to 0$. We compare the expansion coefficients with those
exactly calculated by the diagrammatic perturbative method, and find
some inconsistency. It causes a question in which sense the Wegner--Houghton
equation is really exact. 
\end{abstract}

\section{Introduction}
\label{sec:intro}

The renormalization group (RG) approach, perhaps, is the most extensively used one in numerous
studies of critical phenomena~\cite{Amit,Sornette}. Particularly, the perturbative RG approach 
to the $\varphi^4$ or Ginzburg--Landau model is widely known~\cite{Wilson,Ma,Justin,Kleinert}. 
However, the perturbative approach suffers from some problems~\cite{K1}. Therefore it is interesting to look
for a nonperturbative approach. Historically, nonperturbative RG equations have been developed in parallel
to the perturbative ones. These are so called exact RG equations (ERGE).
The method of deriving such RG equations is close in spirit to the famous Wilson's approach, 
where the basic idea is to integrate out the short--wave fluctuations
corresponding to the wave vectors within $\Lambda/s < q < \Lambda$ with the
upper (or ultraviolet) cutoff parameter $\Lambda$ and the renormalization scale $s>1$.
The oldest nonperturbative equation of this kind, originally presented by 
Wegner and Houghton~\cite{WH}, uses the sharp momentum cutoff. Later, 
a similar equation with smooth momentum cutoff has been proposed by Polchinski~\cite{Polchinski}.
The RG equations of this class are reviewed in~\cite{Report1}.

According to the known classification~\cite{Report1,Report2},
there is another class of nonperturbative RG equations proposed
by~\cite{Wetterich} and reviewed in~\cite{Report2}.
Some relevant discussion can be found in~\cite{Report1}, as well.
Such equations describe the variation of an average effective action 
$\Gamma_k [\phi]$ depending on the running cutoff scale $k$. Here 
$\phi({\bf x}) = \langle \varphi({\bf x}) \rangle$
is the averaged order--parameter field (for simplicity, we refer to the case of scalar field). 
According to~\cite{Wetterich}, the averaging is performed over volume $\sim k^{-d}$
such that the fluctuation degrees of freedom with momenta $q>k$ are effectively integrated
out. In fact, the averaging over volume $\sim k^{-d}$
is the usual block--spin--averaging procedure of the real--space
renormalization. At the same time, the fluctuations with
$q \lesssim k$ are suppressed by a smooth infrared cutoff. 
As one can judge from~\cite{Report2}, the existence of a deterministic relation
between the configuration of external source $\{ J({\bf x}) \}$ and 
that of the averaged order parameter $\{ \phi({\bf x}) \}$
is (implicitly) assumed in the nonperturbative derivation of the RG flow equation.
Namely, it is stated 
(see the text between (2.28) and (2.29) in~\cite{Report2}) that
$\delta J({\bf x}) / \delta \phi({\bf y})$ is the inverse of 
$\delta \phi({\bf x}) / \delta J({\bf y})$,
which has certain meaning as a  matrix identity. 
To make this point clearer, let us consider a toy example $\vec{J} = A \vec{\phi}$, where 
$\vec{J}=\left(J({\bf x}_1),J({\bf x}_2), \ldots, J({\bf x_N}) \right)$ and 
$\vec{\phi}= \left(\phi({\bf x}_1),\phi({\bf x}_2), \ldots, \phi({\bf x_N}) \right)$ are $N$--component
vectors and $A$ is a matrix of size $N \times N$. In this case 
$\partial J({\bf x}_i) / \partial \phi({\bf x}_j)$ is the element $A_{ij}$ of matrix $A$,
whereas $\partial \phi({\bf x}_i) / \partial J({\bf x}_j)$ is the element 
$\left(A^{-1} \right)_{ij}$ of the inverse matrix $A^{-1}$.
In the continuum limit $N \to \infty$, this toy example corresponds to a linear
dependence between $\{ J({\bf x}) \}$ and $\{ \phi({\bf x}) \}$. The calculation of derivative always
implies the linearisation around some point, so that the matrix identity
used in~\cite{Report2} (as a continuum limit in the above example)
has a general meaning. However, it  makes sense only if there exists 
a deterministic relation between the configurations of $\phi({\bf x})$
and $J({\bf x})$ or, in a mathematical notation, if there exist
mappings $f : \{ J({\bf x}) \} \to \{ \phi({\bf x}) \}$ 
and $f^{-1} : \{ \phi({\bf x}) \} \to \{ J({\bf x}) \}$. 
On the other hand, according to the block--averaging, the values of $\phi({\bf x})$ should be
understood as the block--averages. These, of course, are not uniquely
determined by the external sources, but are fluctuating quantities.
So, we are quite sceptical about the exactness of such an approach of averaged effective action.

The integration over fluctuation degrees with momenta $q>k$ does not alter the
behavior of the infrared modes, directly related to the critical exponents. 
From this point of view, the approach based on the equations of Wegner--Houghton 
and Polchinski type seems to be more natural. 
These are widely believed to be the exact 
RG equations, although, in view of our currently presented results, it turns out to be 
questionable in which sense they are really exact. In any case, the
nonperturbative RG equations
cannot be solved exactly, therefore a suitable truncation is used. The convergence
of several truncation schemes and of the derivative expansion
has been widely studied in~\cite{Aoki,Aoki2,Morris,Morris2,CDMV}. Here~\cite{CDMV}
refers to the specific approach of~\cite{Wetterich}. 
A review about all the methods of approximate solution can be found in~\cite{Delamotte}.

Another problem is to test and verify the nonperturbative RG equations,
comparing the results with the known exact and rigorous solutions,
as well as with the results of the perturbation theory.
In~\cite{Morris}, the derivative expansion of the RG $\beta$--function has been considered,
showing the agreement up to the second order between the perturbative results and those 
obtained from the Legendre flow equation, which also belongs to the same class of RG equations
as the Wegner--Houghton and Polchinski equations.
It has been stated in~\cite{Aoki} that the critical exponent $\nu$, extracted from the Wegner--Houghton 
equation in the local potential approximation, agrees with the $\varepsilon$--expansion
up to the $O(\varepsilon)$ order, as well as with the $1/n$ 
($1/N$ in the notations of~\cite{Aoki})
expansion in the leading order.
However, looking carefully on the results of~\cite{Aoki}, one should make clear
that ``the leading order of the $1/n$ expansion" in this case is no more than
the zeroth order, whereas the expansion coefficient at $1/n$ 
is inconsistent with that proposed by the perturbative RG calculation at 
any fixed dimension $d$ except only $d=4$. 
The inconsistency 
could be understood from the point of view that the Wegner--Houghton equation
has been solved approximately. Therefore it would be interesting to verify
whether the problem is eliminated beyond the local potential approximation.
One should also take into account that the perturbative
RG theory is not rigorous and, therefore, we think that a possible
inconsistency still would not prove that something is really wrong
with the nonperturbative RG equation. 
In any case, it is a remarkable fact that correct RG eigenvalue spectrum and critical
exponents are obtained in the local potential approximation at $n \to \infty$
from the Wegner--Houghton equation~\cite{Aoki}, as well as from similar RG
equations~\cite{Morris3},
in agreement with the known exact and rigorous results for the spherical model.
It shows that some solutions, being not exact, nevertheless can lead to exact critical exponents.
From this point of view, it seems also possible that some kind of approximations,
made in the derivation of an RG equation, are not harmful for the critical exponents.

We propose a simple test of the Wegner--Houghton
equation: to verify the expansion of the renormalized action $S$ of the $\varphi^4$ 
model in powers of the coupling constant $u$ at $u \to 0$  in the high--temperature phase. 
Such a test is rigorous, in the sense that the natural domain of validity of the perturbation theory
is considered. 
We think that it would be quite natural to start with such a relatively simple and straightforward
test before passing to more complicated ones, considered in~\cite{Aoki,Morris,Morris3}.
We have made this simplest test in our paper and have found that the
Wegner--Houghton equation fails to give all correct expansion coefficients.
We have also proposed another derivation
of the Wegner--Houghton equation (Secs.~\ref{sec:elementary},~\ref{sec:WH}). 
It is helpful to clarify the origin of the mentioned inconsistency.
It is also less obscure from the point of view that the 
used assumptions and approximations are clearly stated. As regards the derivation in~\cite{WH},
at least one essential step is obscure and apparently contains an implicit 
approximation which, in very essence, is analogous to that pointed out in our derivation.
We will discuss this point in Sec.~\ref{sec:WH}.

\section{An elementary step of renormalization}
\label{sec:elementary}

 To derive a nonperturbative RG equation for the $\varphi^4$ model, we should start with
some elementary steps, as explained in this section.
 
Consider the action $S[\varphi]$ which depends on the configuration of the
order parameter field $\varphi({\bf x})$ depending on coordinate ${\bf x}$.
By definition, it is related to the Hamiltonian $H$ of the model via $S  = H/T$, 
where $T$ is the temperature measured in energy units. In general, $\varphi({\bf x})$ 
is an $n$--component vector with components $\varphi_j({\bf x})$ given in the Fourier 
representation as $\varphi_j({\bf x}) = V^{-1/2} \sum_{k<\Lambda} \varphi_{j,{\bf k}} e^{i{\bf kx}}$,
where $V=L^d$ is the volume of the system, $d$ is the spatial dimensionality, 
and $\Lambda$ is the upper cutoff of the wave vectors. We consider the action of the
Ginzburg--Landau form. For simplicity, we include only the $\varphi^2$ and $\varphi^4$
terms. The action of such $\varphi^4$ model is given by
\begin{equation}
S[\varphi] = \sum\limits_{j,{\bf k}} \Theta({\bf k}) \varphi_{j,{\bf k}} \varphi_{j,-{\bf k}}
+ u V^{-1} \sum\limits_{j,l,{\bf k}_1,{\bf k}_2,{\bf k}_3}
\varphi_{j,{\bf k}_1} \varphi_{j,{\bf k}_2} \varphi_{l,{\bf k}_3}
\varphi_{l,-{\bf k}_1 -{\bf k}_2 -{\bf k}_3} \;,
\label{eq:action}
\end{equation}
where $\Theta({\bf k})$ is some function of wave vector ${\bf k}$, e.~g.,
$\Theta({\bf k})=r_0+ck^2$ like in theories of critical phenomena~\cite{Ma,Justin,Kleinert,K1}.
In the sums we set $\varphi_{l,{\bf k}} =0$ for $k>\Lambda$. 

The renormalization group (RG) transformation implies the integration over $\varphi_{j,{\bf k}}$
for some set of wave vectors with $\Lambda' < k < \Lambda$, i.~e., 
the Kadanoff's transformation, followed by certain rescaling procedure~\cite{Ma}.
The action under the Kadanoff's transformation is changed from $S[\varphi]$ to 
$S_{\mathrm{tra}}[\varphi]$ according to the equation
\begin{equation}
e^{-S_{\mathrm{tra}}[\varphi]} = 
\int e^{-S[\varphi]} \prod\limits_{j,\Lambda'<k < \Lambda} d \varphi_{j,{\bf k}} \;.
\label{eq:Kd}
\end{equation}
Alternatively, one often writes $-S_{\mathrm{tra}}[\varphi]+A L^d$ instead of 
$-S_{\mathrm{tra}}[\varphi]$ to separate the constant
part of the action $A L^d$. This, however, is merely a redefinition of $S_{\mathrm{tra}}$, and for our
purposes it is suitable to use~(\ref{eq:Kd}). 
Note that $\varphi_{j,{\bf k}}=\varphi'_{j,{\bf k}} + i \varphi''_{j,{\bf k}}$ 
is a complex number and $\varphi_{j,-{\bf k}} = \varphi^*_{j,{\bf k}}$
holds (since $\varphi_j({\bf x})$ is always real), so that the integration over 
$\varphi_{j,{\bf k}}$ means in fact the integration over real and imaginary parts of $\varphi_{j,{\bf k}}$
for each pair of conjugated wave vectors ${\bf k}$ and $-{\bf k}$.

The Kadanoff's transformation~(\ref{eq:Kd}) can be split in a sequence of elementary steps
$S[\varphi] \to S_{\mathrm{tra}}[\varphi]$ of the repeated integration given by 
\begin{equation}
e^{-S_{\mathrm{tra}}[\varphi]} = 
\int\limits_{-\infty}^{\infty} \int\limits_{-\infty}^{\infty} 
e^{-S[\varphi]} d \varphi'_{j,{\bf q}} d \varphi''_{j,{\bf q}} 
\label{eq:element}
\end{equation}
for each $j$ and ${\bf q} \in \Omega$, where $\Omega$ is the subset of independent wave vectors
($\pm {\bf q}$ represent one independent mode) within $\Lambda' < q < \Lambda$.
Thus, in the first elementary step of renormalization we have to insert the original
action~(\ref{eq:action}) into~(\ref{eq:element}) and perform the integration for one
chosen $j$ and ${\bf q} \in \Omega$. In an exact treatment we must take into account
that the action is already changed in the following elementary steps. 

For $\Lambda'> \Lambda/3$,
we can use the following exact decomposition of~(\ref{eq:action}) 
\begin{equation}
S[\varphi] = A_0 + A_1 \varphi_{j,{\bf q}} + A_1^* \varphi_{j,-{\bf q}}
+ A_2 \varphi_{j,{\bf q}} \varphi_{j,-{\bf q}} + B_2 \varphi^2_{j,{\bf q}}
+ B_2^* \varphi^2_{j,-{\bf q}} + A_4  \varphi^2_{j,{\bf q}} \varphi^2_{j,-{\bf q}} \;,
\label{eq:dc}
\end{equation}
where
\begin{eqnarray}
A_0 &=& \left. S \right|_{\varphi_{j,\pm {\bf q}}=0}  \;, \\
A_1 &=& \left. \frac{\partial S}{\partial \varphi_{j,{\bf q}}} \right|_{\varphi_{j,\pm {\bf q}}=0}
= 4u V^{-1} \sum'\limits_{l,{\bf k}_1,{\bf k}_2} 
\varphi_{j,{\bf k}_1} \varphi_{l,{\bf k}_2} \varphi_{l,-{\bf q} -{\bf k}_1 -{\bf k}_2} \;, \label{eq:A1} \\
A_2 &=& \left. \frac{\partial^2 S}{\partial \varphi_{j,{\bf q}} \partial \varphi_{j,-{\bf q}}}
\right|_{\varphi_{j,\pm {\bf q}}=0}
= \Theta({\bf q}) + \Theta(-{\bf q}) + 4u V^{-1} \sum'\limits_{l,{\bf k}} (1+ 2 \delta_{lj})
\mid \varphi_{l,{\bf k}} \mid^2 \;, \label{eq:A2} \\
B_2 &=& \frac{1}{2} \left. \frac{\partial^2 S}{\partial \varphi_{j,{\bf q}}^2} 
\right|_{\varphi_{j,\pm {\bf q}}=0}
= 2u V^{-1} \sum'\limits_{l,{\bf k}} (1+ 2 \delta_{lj})
\varphi_{l,{\bf k}} \varphi_{l,-2 {\bf q}- {\bf k}} \;, \\
A_4 &=& \left. \frac{1}{4} \frac{\partial^4 S}{\partial^2 \varphi_{j,{\bf q}} \partial^2 \varphi_{j,-{\bf q}}}
\right|_{\varphi_{j,\pm {\bf q}}=0} = 6u V^{-1} \;.
\end{eqnarray}
Here the sums are marked by a prime to indicate that terms containing $\varphi_{j,\pm {\bf q}}$ are omitted.
This is simply a splitting of~(\ref{eq:action}) into parts with all possible powers of $\varphi_{j,\pm {\bf q}}$. 
The condition $\Lambda'> \Lambda/3$, as well as the existence of the upper
cutoff for the wave vectors, 
ensures that terms of the third power are absent in~(\ref{eq:dc}).
Besides, the derivation is performed formally considering all $\varphi_{l,{\bf k}}$
as independent variables.

Taking into account~(\ref{eq:dc}), as well as the fact that $A_1 =A_1'+i A_1''$
and $B_2 =B_2'+i B_2''$ are complex numbers, the transformed action after the first elementary 
renormalization step reads
\begin{eqnarray}
&&\hspace{-2ex} S_{\mathrm{tra}}[\varphi] = A_0 - \ln \left\{ \int\limits_{-\infty}^{\infty} \int\limits_{-\infty}^{\infty}
\exp \left[ -2 \left( A_1' \varphi'_{j,{\bf q}} - A_1'' \varphi''_{j,{\bf q}} \right) \right] 
\, \exp \left[ -A_2 \left( {\varphi'_{j,{\bf q}}}^2 + {\varphi''_{j,{\bf q}}}^2 \right) \right] \right.
\nonumber \\
&&\hspace{21ex} \times \exp \left[ -2 B_2' \left( {\varphi'_{j,{\bf q}}}^2 - {\varphi''_{j,{\bf q}}}^2 \right)
+ 4 B_2'' \varphi'_{j,{\bf q}} \varphi''_{j,{\bf q}} \right] \nonumber \\ 
&&\left. \hspace{21ex} \times \exp \left[ -A_4 \left( {\varphi'_{j,{\bf q}}}^2 
+ {\varphi''_{j,{\bf q}}}^2 \right)^2 \right]  d \varphi'_{j,{\bf q}} d \varphi''_{j,{\bf q}} \right\} \;.   
\label{eq:S1}
\end{eqnarray}

Considering only the field configurations which are relevant in the thermodynamic limit $V \to \infty$,
Eq.~(\ref{eq:S1}) can be simplified, omitting the terms with $B_2$ and $A_4$. Really, using the coordinate
representation $\varphi_{l,{\bf k}} = V^{-1/2} \int   \varphi_l({\bf x}) \, e^{-i {\bf kx}} d{\bf x}$, we can write
\begin{equation}
B_2 = 2u V^{-1} \left\{ \left[ \sum\limits_{l,j} (1+2 \delta_{lj} ) 
\int \varphi_l^2 ({\bf x}) e^{i{\bf qx}} \, d{\bf x} \right] -3 \varphi_{j,-{\bf q}}^2 \right\} \;.
\end{equation} 
The quantity $V^{-1} \int \varphi_l^2 ({\bf x}) e^{i{\bf qx}} \, d {\bf x}$ is an average
of $\varphi_l^2 ({\bf x})$ over the volume with oscillating weight factor $e^{i{\bf qx}}$.
This quantity vanishes for relevant configurations in the thermodynamic limit: due to
the oscillations, positive and negative contributions are similar in magnitude and cancel at $V \to \infty$. 
Since $\langle \mid \varphi_{j,-{\bf q}} \mid^2 \rangle = 
\langle \mid \varphi_{j,{\bf q}} \mid^2 \rangle$ is the Fourier transform of the two--point correlation
function, it is bounded at $V \to \infty$ and, hence, $\varphi_{j,-{\bf q}}^2$ also is bounded for
relevant configurations giving nonvanishing contribution to the statistical averages $\langle \cdot \rangle$
in the thermodynamic limit. Consequently, for these configurations, $A_2$ is a quantity of order $\mathcal{O}(1)$,
whereas $V^{-1} \varphi_{j,-{\bf q}}^2$ and $B_2$ vanish at $V \to \infty$. Note, however,
that the term with $A_4= \mathcal{O} \left( V^{-1} \right)$ in~(\ref{eq:S1}) 
cannot be neglected unless $A_2$ is positive. One can judge that the latter condition is satisfied for 
the relevant field configurations due to existence of the thermodynamic limit for the RG flow.

Omitting the terms with $B_2$ and $A_4$, the integrals in~(\ref{eq:S1}) can be easily calculated. It yields
\begin{equation}
S_{\mathrm{tra}} [\varphi] = S'[\varphi] + \Delta S_{\mathrm{tra}}^{el} [\varphi] \;, 
\end{equation}
where $S'[\varphi]=A_0$ is the original action, where only the $\pm {\bf q}$ modes of the $j$-th field component 
are omitted, whereas $\Delta S_{\mathrm{tra}}^{el} [\varphi]$ represents the elementary variation of 
the action given by
\begin{equation}
\Delta S_{\mathrm{tra}}^{el} [\varphi] = \ln \left( \frac{A_2}{\pi} \right) - \frac{ \mid A_1 \mid^2}{A_2}  \;.
\label{eq:dS1}
\end{equation}
According to the arguments provided above, this equation is exact for the relevant 
field configurations with $A_2>0$ in the thermodynamic limit.

The contributions to~(\ref{eq:A1}) and~(\ref{eq:A2}) provided by modes with wave vectors ${\bf k}$, obeying two
relations $\Lambda- d \Lambda < k < \Lambda$ and ${\bf k} \ne \pm {\bf q}$,
are irrelevant in the thermodynamic limit at $d \Lambda \to 0$. 
It can be verified by the method of analysis introduced in Sec.~\ref{sec:elementary}.
Hence, Eq.~(\ref{eq:dS1}) can be written as 
\begin{equation}
\Delta S_{\mathrm{tra}}^{el} [\varphi] = \ln \left( \frac{\widetilde{A}_2}{\pi} \right) 
- \frac{ \mid \widetilde{A}_1 \mid^2}{\widetilde{A}_2}  + \delta S_{\mathrm{tra}}^{el} [\varphi] \;,
\label{eq:dS1x}
\end{equation}
where 
\begin{eqnarray}
\widetilde{A}_1 &=& P \frac{\partial S}{\partial \varphi_{j,{\bf q}}}   
= 4u V^{-1} \sum\limits_{l,{\bf k}_1,{\bf k}_2}^{\Lambda - d \Lambda} 
\varphi_{j,{\bf k}_1} \varphi_{l,{\bf k}_2} \varphi_{l,-{\bf q} -{\bf k}_1 -{\bf k}_2} \;, \label{eq:A1x} \\
\widetilde{A}_2 &=& P \frac{\partial^2 S}{\partial \varphi_{j,{\bf q}} \partial \varphi_{j,-{\bf q}}}
= \Theta({\bf q}) + \Theta(-{\bf q}) + 4u V^{-1} \sum\limits_{l,{\bf k}}^{\Lambda - d \Lambda} (1+ 2 \delta_{lj})
\mid \varphi_{l,{\bf k}} \mid^2 \;, \label{eq:A2x} 
\end{eqnarray}
and $\delta S_{\mathrm{tra}}^{el} [\varphi]$ is a vanishingly small correction
in the considered limit.
Here the operators $P$ set to zero all $\varphi_{j,{\bf k}}$ 
within the shell $\Lambda - d \Lambda < k < \Lambda$ (i.~e., the derivatives are 
evaluated at zero $\varphi_{j,{\bf k}}$ for ${\bf k}$ within the shell), 
and the upper border $\Lambda - d \Lambda$ for sums implies that we set $\varphi_{l,{\bf k}}=0$
for $k> \Lambda-d \Lambda$. 
 The above replacements are meaningful, since they allow to obtain easily the
Wegner--Houghton equation, as discussed in the following section.

\section{Superposition hypothesis and the \\ Wegner--Houghton equation}
\label{sec:WH}

Intuitively, it could seem very reasonable that the result of integration over Fourier modes 
within the shell $\Lambda-d \Lambda<k<\Lambda$ at $d \Lambda \to 0$ can be represented as a 
superposition of elementary contributions given by~(\ref{eq:dS1x}), neglecting the
irrelevant corrections $\delta S_{\mathrm{tra}}^{el} [\varphi]$.
We will call this idea the superposition hypothesis. 

We remind, however, that strictly exact treatment requires a sequential integration 
of $\exp(-S[\varphi])$ over a set of $\varphi_{j,{\bf q}}$. 
The renormalized action changes after each such integration, and these changes influence
the following steps. 
A problem is to estimate the discrepancy between the results of two methods:
(1) the exact integration and (2) the superposition approximation.
Since it is necessary to perform infinitely many integration steps
in the thermodynamic limit, the problem is nontrivial and the
superposition hypothesis cannot be rigorously justified. 

Nevertheless, the summation of elementary contributions in accordance with the superposition
hypothesis leads to the known Wegner--Houghton equation~\cite{WH}. In this case the variation
of the action due to the integration over shell reads
\begin{equation}
\Delta S_{\mathrm{tra}}[\varphi] = \frac{1}{2} \sum\limits_j 
\sum\limits_{\Lambda-d \Lambda <q < \Lambda}
\left[ \ln \left( \frac{\widetilde{A}_2(j,{\bf q})}{\pi} \right) 
- \frac{ \mid \widetilde{A}_1(j,{\bf q}) \mid^2}{\widetilde{A}_2(j,{\bf q})} \right] \;.
\label{eq:WH}
\end{equation}
It is exactly consistent with Eq.~(2.13) in~\cite{WH}.
The factor $1/2$ appears, since only half of the wave vectors represent independent modes.
Here we have indicated that the quantities $\widetilde{A}_1$ and $\widetilde{A}_2$ depend on the current 
$j$ and ${\bf q}$. They depend also on the considered field configuration
$[\varphi]$. If $\widetilde{A}_1$ and $\widetilde{A}_2$ are represented by the derivatives of $S[\varphi]$
(see~(\ref{eq:A1x}) and~(\ref{eq:A2x})), then the equation is written exactly as in~\cite{WH}.

To avoid possible confusion, one has to make clear that the operators 
$P$ in~(\ref{eq:A1x}) and~(\ref{eq:A2x}) influence
the result, as discussed further on. It means that the equation 
where these operators are simply omitted, referred 
in the review paper~\cite{Report1} as the Wegner--Houghton
equation, is not really the Wegner--Houghton equation.

The derivation in~\cite{WH} is somewhat different. Instead of performing only one
elementary step of integration first, the expansion of Hamiltonian in terms of all shell variables
is made there. The basic method of~\cite{WH} is to show that, in the thermodynamic
limit at $d \Lambda \to 0$, the expansion consists of terms containing no more than 
two derivatives with respect to the field components. Moreover, it is assumed 
implicitly that only the diagonal terms
with ${\bf k' = -k}$ are important finally, when performing the
summation over the wave vectors ${\bf k, k'}$. It leads to Eq.~(2.12) in~\cite{WH}. The omitting
of nondiagonal terms is equivalent to the superposition assumption we 
discussed already. Indeed, in this and only in this case the integration
over the shell variables can be performed independently, as if the superposition
principle were hold. Hence, essentially the same approximation is used in~\cite{WH}
as in our derivation, although it is not stated explicitly.

Our derivation refers to the $\varphi^4$ model, whereas in the form with derivatives 
the equation may have a more general validity, as supposed in~\cite{WH}. Indeed, (\ref{eq:dS1x})
remains correct for a generalized model provided that higher than second order derivatives of $S[\varphi]$
vanish for relevant field configurations in the thermodynamic limit. It, in fact, has
been assumed and shown in~\cite{WH}.
Based on similar arguments we have used already, the latter assumption can be justified
for certain class of models, for which the action is represented by a linear combination
of $\varphi^m$--kind  terms with wave--vector dependent weights and vanishing sum of 
the wave vectors $\sum_{l=1}^m {\bf k}_l = {\bf 0}$ related to the $\varphi$ factors.
In this case we have
\begin{eqnarray}
\widetilde{A}_1(j,{\bf q}) &=& \frac{\partial S}{\partial \varphi_{j,{\bf q}}} -
\varphi_{j,-{\bf q}} \, \frac{\partial^2 S}{\partial \varphi_{j,{\bf q}} \varphi_{j,-{\bf q}}} 
\label{eq:A1d} \;, \\
\widetilde{A}_2(j,{\bf q}) &=& \frac{\partial^2 S}{\partial \varphi_{j,{\bf q}} \varphi_{j,-{\bf q}}} 
\end{eqnarray}
for the relevant configurations at $V \to \infty$ and $d \Lambda \to 0$.
The second term in~(\ref{eq:A1d}) appears because the derivative $\partial S / \partial \varphi_{j,{\bf q}}$
contains relevant terms with $\varphi_{j,-{\bf q}}$, which have to be removed. 
The influence of the operators $P$ is seen from~(\ref{eq:A1x}) and~(\ref{eq:A1d}).

Here we do not include the second, i.~e., the rescaling
step of the RG transformation. It, however, can be easily calculated for any given action,
as described, e.~g., in~\cite{Ma}. It is not relevant four our further considerations.

\section{The weak coupling limit}
\label{sec:expansion}

Here we consider
the weak coupling limit $u \to 0$  of the model with $\Theta({\bf k}) = r_0 + ck^2$ at 
a given positive $r_0$, i.~e., in the high temperature phase. 
In this case $\Delta S_{\mathrm{tra}}[\varphi]$ can be expanded in powers
of $u$. It is the natural domain of validity of the perturbation
theory, and the expansion coefficients can be calculated exactly by the known 
methods applying the Feynman diagram technique and the 
Wick's theorem~\cite{Ma,Justin,Report2}. 
On the other hand, the expansion can be performed in~(\ref{eq:WH}). 
Our aim is to compare the results of both methods to check the correctness of~(\ref{eq:WH}),
since the latter equation is based on assumptions.

Let us denote by $\Delta \tilde S_{\mathrm{tra}}[\varphi]$ the variation of $S[\varphi]$ omitting
the constant (independent of the field configuration) part. Then the expansion in powers of $u$ reads
\begin{equation}
\Delta \tilde S_{\mathrm{tra}}[\varphi] = \Delta S_1[\varphi] \, u
+ \left( \Delta S_2^{(a)}[\varphi] + \Delta S_2^{(b)}[\varphi] + \Delta S_2^{(c)}[\varphi] \right)
u^2 + \mathcal{O} \left( u^3 \right) \;,
\end{equation}
where the expansion coefficient at $u^2$ is split in three parts $\Delta S_2^{(a)}[\varphi]$, 
$\Delta S_2^{(b)}[\varphi]$, and $\Delta S_2^{(c)}[\varphi]$ corresponding to the $\varphi^2$,
$\varphi^4$, and $\varphi^6$ contributions, respectively. 
The contribution of order $u$ is related to the diagram \mbox{\vloop,} whereas the three second--order
contributions --- to the diagrams \mbox{\sunset,} \mbox{\vvertice,} and \mbox{\six.}
The diagram technique represents the expansion of $-S[\varphi]$ in terms
of connected Feynman diagrams, where the coupled lines are associated with the 
Gaussian averages. In particular, the Fourier transformed two--point correlation function in 
the Gaussian approximation
$G_0({\bf k})=\langle \varphi_{j,{\bf k}} \varphi_{j,-{\bf k}} \rangle_0=1/[2 \Theta({\bf k})]$ 
appears due to the integration over $\varphi'_{j,{\bf k}}$ and $\varphi''_{j,{\bf k}}$.
It is represented as the coupling of lines, in such a way that each line related to the wave vector 
${\bf k}$ and vector--component $j$ is coupled with another line having the 
wave vector $-{\bf k}$ and the same component $j$. Thus, if we integrate
over $\varphi_{j,{\bf k}}$ within $\Lambda - d \Lambda < k < \Lambda$
in~(\ref{eq:Kd}), then it corresponds to the coupling of lines in the same 
range of wave vectors in the diagram technique. According
to the Wick's theorem, one has to sum over all possible couplings,
which finally yields the summation (integration) over the wave vectors
obeying the constraint $\Lambda -d \Lambda < k < \Lambda$
for each of the coupled lines associated with the factors
$G_0({\bf k})$. In the $n$--component case, it is suitable to represent the $\varphi^4$ vertex
as \dshv, where the same index $j$ is associated with two solid lines
connected to one node. The above diagrams are given by the sum of all possible couplings of the
vertices \mbox{\dshv,} yielding the corresponding topological pictures when the dashed lines
shrink to points. In this case factor $n$ corresponds to each closed loop of solid lines,
which comes from the summation over $j$. For a complete definition of the diagram technique,
one has to mention that factors $-uV^{-1}$ are related to the dashed lines,
$G_0({\bf k})$ -- to the coupled solid lines, and the 
fields $\varphi_{j,{\bf k}}$ -- to the outer uncoupled solid lines. Besides, each diagram
contains a combinatorial factor. For a diagram consisting of $m$ vertices
\mbox{\dshv,} it is the number of all possible couplings of (numbered) lines, divided by $m!$.

At $d \Lambda \to 0$, the diagrammatic calculation  for the $n$--component case yields
\begin{eqnarray}
\Delta S_1[\varphi] &=& \frac{K_d \Lambda^{d-1}}{\Theta(\Lambda)} \, (n+2) \, d \Lambda
\sum\limits_{j,{\bf k}}^{\Lambda -d \Lambda} \mid \varphi_{j,{\bf k}} \mid^2 \label{eq:viens} \\
\Delta S_2^{(b)} [\varphi] &=& -4 V^{-1} \sum\limits_{j,l,{\bf k}_1,{\bf k}_2,{\bf k}_3}^{\Lambda-d \Lambda}
\varphi_{j,{\bf k}_1} \varphi_{j,{\bf k}_2} \varphi_{l,{\bf k}_3}
\varphi_{l,-{\bf k}_1 -{\bf k}_2 -{\bf k}_3} \label{eq:b} \\
&& \hspace{15ex} \times \left[ (n+4) Q \left( {\bf k}_1+{\bf k}_2, \Lambda, d \Lambda \right)
+4 Q \left( {\bf k}_1+{\bf k}_3, \Lambda, d \Lambda \right) \right] \nonumber \\
\Delta S_2^{(c)} [\varphi] &=& -8 V^{-2} \sum\limits_{i,j,l,{\bf k}_1,{\bf k}_2,{\bf k}_3,
{\bf k}_4,{\bf k}_5}^{\Lambda-d \Lambda}
\varphi_{i,{\bf k}_1} \varphi_{i,{\bf k}_2} \varphi_{j,{\bf k}_3} \varphi_{j,{\bf k}_4} \varphi_{l,{\bf k}_5}
\varphi_{l,-{\bf k}_1 -{\bf k}_2 -{\bf k}_3-{\bf k}_4 -{\bf k}_5} \nonumber \\
&& \hspace{15ex} \times G_0 \left({\bf k}_1+{\bf k}_2+{\bf k}_3 \right)
\, \mathcal{F} \left( \mid {\bf k}_1+{\bf k}_2+{\bf k}_3 \mid,\Lambda,d \Lambda \right) \label{eq:c} \;,
\end{eqnarray}
where $K_d=S(d)/(2 \pi)^d$, $S(d)=2 \pi^{d/2}/ \Gamma(d/2)$ is the area of unit sphere in $d$ dimensions,
$\Theta(\Lambda)$ is the value of $\Theta({\bf k})$ at $k=\Lambda$,
whereas $\mathcal{F}(k,\Lambda,d \Lambda)$ is a cutoff function which has the value
$1$ within $\Lambda- d \Lambda < k < \Lambda$ and zero otherwise. The quantity $Q$ is given by
\begin{equation}
Q({\bf k},\Lambda,d \Lambda) = V^{-1} \sum\limits_{\Lambda- d \Lambda < q < \Lambda} 
G_0({\bf q}) \, G_0({\bf k}-{\bf q}) \, \mathcal{F}( \mid {\bf k}-{\bf q} \mid,\Lambda,d \Lambda)  \;.
\label{eq:QQ}
\end{equation}

Below we will give some details of calculation of~(\ref{eq:b}), which is the
most important term in our further discussion.
To obtain this result, we have dechipered the \vvertice diagram as 
a sum of three diagrams of different topologies made of vertices 
\mbox{\dshv,} i.~e., \mbox{\dshvv, \dshvvv, and \dshvvvv,}
providing the same topological picture \vvertice when shrinking the dashed lines 
to points. Recall that any loop made of solid lines of \dshv gives a factor $n$,
and one needs also to compute the combinatorial factors. 
For the above three diagrams, the resulting factors are $4n$, $16$, and 
$16$, which enter the prefactors of $Q$ in~(\ref{eq:b}). 
To obtain the correct sign, we recall that the diagram expansion is for
$-S[\varphi]$. The other diagrams are calculated in a similar way.

The expansion of~(\ref{eq:WH}) gives no contribution $\Delta S_2^{(a)} [\varphi]$,
and we have skipped it in the diagrammatic calculation as an irrelevant
term, which vanishes faster than $\propto d \Lambda$ at $d \Lambda \to 0$
in the thermodynamic limit $V \to \infty$.
The expansion of the logarithm term in~(\ref{eq:WH}) yields $\Delta S_1 [\varphi]$
exactly consistent with~(\ref{eq:viens}). Similarly, 
$\Delta S_2^{(c)} [\varphi]$ is exactly consistent with~(\ref{eq:c}). 

One has to remark that two propogators are involved in~(\ref{eq:QQ}) and, therefore,
the volume of summation region with nonvanishing cut function $\mathcal{F}$
shrinks as $(d \Lambda)^2$ for a given nonzero wave vector ${\bf k}$ at $d \Lambda \to 0$.
However, there is a contribution linear in $d \Lambda$
for ${\bf k = 0}$. As a result, a contribution 
proportional to $d \Lambda$ appears in~(\ref{eq:b}).

Note that the contributions~(\ref{eq:viens}) and~(\ref{eq:c}) come
from diagrams with only one coupled line. The term~(\ref{eq:b}) is related to the diagram
with two coupled lines. The expansion of~(\ref{eq:WH})
provides a different result for the corresponding part of $\Delta \tilde S_{\mathrm{tra}}[\varphi]$:
\begin{equation}
\Delta S_2^{(b)} [\varphi] = - \frac{K_d \, \Lambda^{d-1} \, d \Lambda}{ \Theta^2(\Lambda)}   
\; V^{-1} \sum\limits_{j,l,{\bf k}_1,{\bf k}_2}^{\Lambda-d \Lambda}
\left(n+4+4 \delta_{jl} \right)
\mid \varphi_{j,{\bf k}_1} \mid^2 \, \mid \varphi_{l,{\bf k}_2} \mid^2 \;. 
\label{eq:b2}
\end{equation}
Note that~(\ref{eq:b2}) comes from the $\ln \widetilde{A}_2$ term in~(\ref{eq:WH}),
and the calculation is particularly simple 
in this case, since the related sum in~(\ref{eq:A2x}) is independent of ${\bf q}$.
Eq.~(\ref{eq:b2}) is obtained if we set 
$Q({\bf k},\Lambda,d \Lambda) \to \delta_{{\bf k},{\bf 0}} \, Q({\bf 0},\Lambda,d \Lambda)$
in~(\ref{eq:b}) (in this case only the diagonal terms $j=l$ are
relevant when summing up the contributions with
$Q \left( {\bf k}_1+{\bf k}_3, \Lambda, d \Lambda \right)$, as it can be shown by
an analysis of relevant real--space configurations, since
$\langle \varphi_j({\bf x}) \varphi_l({\bf x}) \rangle=0$ holds for $j \ne l$). 
It means that a subset of terms is missing in~(\ref{eq:b2}), as compared to~(\ref{eq:b}).
The following analysis will show that this discrepancy between~(\ref{eq:b}) and~(\ref{eq:b2})
is important.
 
It is interesting to mention that~(\ref{eq:b2}) is obtained also by the diagrammatic perturbation
method if we first integrate out only the mode with $\varphi_{j,\pm {\bf q}}$ and then formally
apply the superposition hypothesis, as in the derivation of the Wegner--Houghton equation.
It shows that the discrepancy between~(\ref{eq:b2}) and~(\ref{eq:b}) arises because in one 
case the superposition hypothesis is applied, whereas in the other
case it is not used.

 The difference between~(\ref{eq:b}) and~(\ref{eq:b2}) can be better seen in the coordinate
representation. In this case (\ref{eq:b}) reads
\begin{eqnarray}
\Delta S_2^{(b)} [\varphi] = &-& (4n+16) \int\int \varphi^2({\bf x}_1) \, R^2({\bf x}_1-{\bf x}_2)
\, \varphi^2({\bf x}_2) \, d {\bf x}_1 d {\bf x}_2 \label{eq:x} \\
&-& 16 \sum\limits_{j,l} \int\int \varphi_j({\bf x}_1) \varphi_l({\bf x}_1) \, R^2({\bf x}_1-{\bf x}_2)
\, \varphi_j({\bf x}_2) \varphi_l({\bf x}_2) \, d {\bf x}_1 d {\bf x}_2 \;, \nonumber
\end{eqnarray}   
where
\begin{equation}
R({\bf x}) = V^{-1} \sum\limits_{\bf q} G_0({\bf q}) \, \mathcal{F}(q,\Lambda,d \Lambda) e^{i{\bf qx}}
\end{equation}
is the Fourier transform of $G_0 \mathcal{F}$, and $\varphi^2({\bf x})=\sum_l \varphi_l^2({\bf x})$.
In three dimensions we have
\begin{equation}
R({\bf x}) =  \frac{\Lambda \, d \Lambda}{(2 \pi)^2 \Theta(\Lambda)} \; \frac{1}{x} \sin(\Lambda x)
\label{eq:cont}
\end{equation}
for any given ${\bf x}$ at $d \Lambda \to 0$ and $L \to \infty$, where $L$ is the linear system size. 
The continuum approximation~(\ref{eq:cont}), however, is not correct for $x \sim L$ and therefore, 
probably, should not be used for the evaluation of~(\ref{eq:x}).

The coordinate representation of~(\ref{eq:b2}) is
\begin{eqnarray}
\Delta S_2^{(b)} [\varphi] &=& - \frac{K_d \, \Lambda^{d-1} \, d \Lambda}{ \Theta^2(\Lambda)}
\, \left[ (n+4) \int\int \varphi^2({\bf x}_1) \, V^{-1}
\, \varphi^2({\bf x}_2) \, d {\bf x}_1 d {\bf x}_2 \right. \nonumber \\
&&\hspace{15ex} \left. +4 \sum\limits_j \int\int \varphi_j^2({\bf x}_1) \, V^{-1}
\, \varphi_j^2({\bf x}_2) \, d {\bf x}_1 d {\bf x}_2 \right] \;.
\label{eq:x2}
\end{eqnarray}
Eq.~(\ref{eq:x2}) represents a relevant contribution at $d \Lambda \to 0$,
as it is proportional to $d \Lambda$. It is obviously not consistent with~(\ref{eq:x}). 
In fact, the term (\ref{eq:x2}) represents 
a mean-field interaction, which is proportional to $1/V$ and independent of the distance,
whereas (\ref{eq:x}) corresponds to another non-local interaction given by $R^2({\bf x}_1-{\bf x}_2)$.
Hence, the Wegner--Houghton equation~(\ref{eq:WH}) does not yield all correct expansion coefficients
at $u \to 0$.

\section{Discussion}
\label{sec:disc}

The results of our test, stated at the end of Sec.~\ref{sec:expansion}, 
reveal some inconsistency between the Wegner--Houghton equation 
and the diagrammatic perturbation theory in the high temperature phase at $u \to 0$.
Since this is the natural domain of validity of the perturbation
theory, there should be no doubts that it produces correct results here, which 
agree with~(\ref{eq:Kd}).
So, the results of our test point to some inconsistency between the 
Wegner--Houghton equation and~(\ref{eq:Kd}), which
causes a question in which sense the Wegner--Houghton equation is really exact.
The same can be asked about the equations of Polchinski type, since these 
(as it is believed) are generalizations of the Wegner--Houghton equation to 
the case of smooth momentum cutoff.
There is no contradiction with the tests of consistency made in~\cite{Aoki,Morris}, 
since our test is independent and quite different.
According to our derivation of the Wegner--Houghton equation and the related discussion,
it turns out that the reason of the inconsistency, likely, is the superposition
approximation (defined at the beginning of Sec.~\ref{sec:WH}) 
used in our paper and analogous approximation implicitly used in~\cite{WH}.
Despite of this problem, the Wegner--Houghton equation is able to reproduce
the exact RG eigenvalue spectrum and critical exponents of the spherical model
at $n \to \infty$~\cite{Aoki}. This fact can be interpreted in such a way
that the superposition approximation (or an analogous approximation) is valid 
to derive such nonperturbative RG equations, which can produce correct 
(exact) critical exponents in some limit cases, at least. 
From a general point of view, it concerns the fundamental question 
about the relation between the form of RG equation and the universal quantities. 
It has been verified in several known studies that the universal quantities are
invariant with respect to some kind of variations in the RG equation, like 
changes in the shape of the momentum cutoff function. This property is known as
the reparametrisation invariance~\cite{Report1}. Probably, the universal
quantities are invariant also with respect to such a variation
of the Wegner--Houghton equation, which makes it exactly consistent with~(\ref{eq:Kd}).
However, this is only a hypothesis.

\section{Conclusions}

\begin{enumerate}

\item
The nonperturbative Wegner--Houghton RG equation has been rederived 
(Secs.~\ref{sec:elementary} and~\ref{sec:WH}), discussing explicitly
some assumptions which are used here. In particular, our derivation assumes the 
superposition of small contributions provided by elementary integration steps over
the short--wave fluctuation modes. We consider it as an approximation.
As discussed in Sec.~\ref{sec:WH}, the original derivation by Wegner and Houghton 
includes essentially the same approximation, although not stated explicitly.
\item
According to our calculation in Sec.~\ref{sec:expansion}, the Wegner--Houghton 
equation is not completely consistent with the diagrammatic
perturbation theory in the limit of small $\varphi^4$ coupling constant $u$ in the
high temperature phase. 
This fact, together with some other important results known from literature,
is discussed in Sec.~\ref{sec:disc}. 
Apart from critical remarks, a hypothesis has been proposed that
the equations of Wegner--Houghton type, perhaps, can give exact  
universal quantities.

\end{enumerate}

\end{document}